\documentclass[conference]{IEEEtran}
\IEEEoverridecommandlockouts
\usepackage{cite}
\usepackage{amsmath,amssymb,amsfonts}
\usepackage{algorithmic}
\usepackage{graphicx}
\usepackage{textcomp}
\usepackage{xcolor}
\usepackage{booktabs} 
\def\BibTeX{{\rm B\kern-.05em{\sc i\kern-.025em b}\kern-.08em
    T\kern-.1667em\lower.7ex\hbox{E}\kern-.125emX}}
\begin{document}

\title{Collective Reasoning Among LLMs: A Framework for Answer Validation Without Ground Truth}

\author{
  \IEEEauthorblockN{ Seyed Pouyan Mousavi Davoudi}
  \IEEEauthorblockA{\textit{Independent Researcher in AI and Statistics} \\
    Tehran, Iran \\
    spouyan.mousavi@gmail.com}
  \and
  \IEEEauthorblockN{ Amin Gholami Davodi}
  \IEEEauthorblockA{\textit{Independent Researcher in AI and Statistics} \\
    \textit{Shahrood University of Technology} \\
    Tehran, Iran \\
    a.g.davodi@gmail.com}
  \and
  \IEEEauthorblockN{ Alireza Amiri Margavi}
  \IEEEauthorblockA{\textit{Computational Modeling and Simulation} \\
    \textit{University of Pittsburgh} \\
    Pittsburgh, USA \\
    ala170@pitt.edu}
  \and

    \IEEEauthorblockN{ Alireza Shafiee Fard}
  \IEEEauthorblockA{\textit{Independent Researcher in AI and Statistics} \\
    \textit{Tehran, Iran} \\
    Shafiee.fard.ar@gmail.com}
  \and
  \IEEEauthorblockN{ Mahdi Jafari}
  \IEEEauthorblockA{\textit{Analytics and Information Science} \\
    \textit{Duquesne University} \\
    Pittsburgh, USA \\
    jafarim@duq.edu}
}

\maketitle

\begin{abstract}
We introduce a new approach in which several advanced large language models—specifically GPT-4-0125-preview, Meta-LLAMA-3-70B-Instruct, Claude-3-Opus, and Gemini-1.5-Flash—collaborate to both produce and answer intricate, doctoral-level probability problems without relying on any single “correct” reference. Rather than depending on an established ground truth, our investigation focuses on how agreement among diverse models can signal the reliability of their outputs and, by extension, reflect the overall quality of the generated questions. To measure this inter-model alignment, we apply a suite of statistical evaluations, including chi-square tests, Fleiss’ Kappa coefficients, and confidence interval calculations, thereby capturing both precision in answers and clarity in question phrasing. Our analysis reveals that Claude and Gemini tend to frame questions more coherently and unambiguously, which is evidenced by their tighter confidence intervals and greater concordance with responding agents. In contrast, LLAMA exhibits wider confidence bands and a lower level of agreement, indicating more variability and reduced consistency in its question formulations. These observations support the notion that a multi-model collaborative strategy not only improves answer dependability but also offers an effective, data-driven mechanism for evaluating and refining question quality when no definitive solution exists. Ultimately, this work delivers actionable insights into enhancing AI-guided reasoning processes through coordinated interactions among heterogeneous language models.
\end{abstract}


\section{Introduction}

Recent breakthroughs in large language models (LLMs) have dramatically altered the landscape of natural language processing, empowering AI systems to perform tasks that once required extensive human expertise. For instance, state-of-the-art architectures—such as OpenAI’s GPT-4, Meta’s LLaMA series, Anthropic’s Claude, Google’s Gemini, and DeepSeek—now routinely generate coherent text, comprehend subtle contextual cues, and solve intricate reasoning problems \cite{guo2025deepseek, touvron2023LLaMA}. These transformative capabilities have unlocked new avenues for automated knowledge creation and validation in domains that traditionally rely on expert human judgment \cite{kojima2022large}.

Nonetheless, a persistent challenge arises when there is no single “correct” answer to gauge output quality. This problem is especially acute in specialized areas like advanced statistics and probability, where manually verifying each model-generated response can be prohibitively slow and expensive \cite{amiri2024enhancing}. In such contexts, conventional approaches—comparing responses against a fixed answer key—often fail, because the dynamic nature of statistical knowledge and the subtleties of probabilistic reasoning cannot be fully captured by static benchmarks \cite{hendrycks2020measuring}. Consequently, new validation strategies that do not depend on predefined ground truths are urgently needed.

One promising direction involves harnessing multiple LLMs simultaneously and combining their strengths to infer a “consensus” answer in the absence of a true label \cite{he2022wisdom}. This approach is analogous to ensemble learning in traditional machine learning, where aggregating predictions from diverse models can improve overall accuracy \cite{dietterich2000ensemble}. It also parallels the “wisdom of crowds” phenomenon in human decision-making, whereby collective judgments often exceed individual expertise \cite{mennis2006wisdom}. By enlisting several LLMs to tackle the same problem and then reconciling their outputs, it becomes possible to approximate correctness even when direct validation is impossible.

Existing research hints that different LLMs exhibit unique reasoning capabilities and biases \cite{taylor2022galactica}, suggesting that a multi-model setup can mitigate individual shortcomings. However, while prior studies have evaluated single-LMM performance \cite{ahn2024large} or experimented with simple ensemble techniques \cite{huang2024enabling}, there remains a significant gap in our understanding of how cutting-edge LLMs might collaborate on complex probabilistic tasks without any ground truth. This gap is particularly pronounced in fields where:
\begin{itemize}
  \item Deep, nuanced reasoning is required to formulate and solve questions.
  \item Manual expert oversight is expensive and does not scale.
  \item Standard automated checks are insufficient due to evolving knowledge.
  \item The rapid advancement of the field makes static answer sets quickly obsolete.
\end{itemize}

To explore these issues, we propose a novel framework grounded in three interrelated theoretical perspectives: collective intelligence, distributed cognition, and consensus formation\footnote{The interplay of these frameworks has been previously explored in cognitive science but is less well-understood in AI collaboration contexts \cite{patrikalakis1999distributed}.}. Instead of analyzing each model in isolation, we examine how multiple LLMs—each contributing its own reasoning strengths—can combine their outputs to produce more reliable answers.

\paragraph{Collective Intelligence} 
Collective intelligence theorizes that a group of diverse agents can outperform any individual member by pooling their varied insights \cite{levy1997collective}. In our setting, each LLM (for example, GPT-4, Claude, LLaMA, and Gemini) brings a different training corpus, architectural bias, and fine-tuning regime to bear on a given problem. By weaving together these distinct perspectives, we aim to reduce the risk of systematic errors that might arise when relying on a single LLM. Empirical evidence suggests that structured consensus mechanisms—such as weighted voting or answer-ranking—can enhance overall solution accuracy\footnote{For instance, weighted ensemble methods have improved predictive stability in other multi-agent systems \cite{woolley2010evidence}.}.

\paragraph{Distributed Cognition} 
Distributed cognition emphasizes how tasks can be split among agents to leverage their complementary capabilities \cite{hutchins1995cognition}. Applied to LLMs, this means assigning subtasks—such as question generation, answer selection, and justification writing—to different models, then synthesizing their outputs. For example, one model might excel at crafting a challenging probabilistic question, while another might provide a more accurate explanation of the solution. By dividing the workload in this manner, we can accumulate a richer set of responses than any single model could generate on its own \cite{du2024large}. Additionally, this framework helps us understand how information and reasoning “flows” between LLMs during collaborative problem-solving \cite{naik2024probabilistic}.

\paragraph{Consensus Formation} 
Models of consensus formation describe how agreement emerges within a group and the conditions necessary for stable convergence \cite{friedkin1990social}. In multi-LLM collaboration, consensus formation concepts guide the design of mechanisms that reconcile disparate answers into a single “consensus” response. Key factors include each model’s influence weight\footnote{In social networks, agent influence can be modeled using adjacency matrices; in LLM ensembles, we analogously assign weights based on model confidence or historical accuracy \cite{olfati2007consensus}.} and the iteration schedule for updating provisional answers. Recent work in multi-agent systems has demonstrated that carefully calibrated consensus rules can significantly reduce disagreement and amplify collective accuracy \cite{baronchelli2018emergence, bahrami2010optimally}.

\paragraph{A Unified Reliability Metric} 
Integrating these three perspectives allows us to formalize a single reliability measure for collaborative LLM outputs. Concretely, we define: R = g(\text{CI},\,\text{DC},\,\text{CF}) where \(\text{CI}\) quantifies the diversity and complementary strengths of the participating models, \(\text{DC}\) captures the effectiveness of task distribution, and \(\text{CF}\) reflects the stability of the consensus protocol. By empirically estimating each component (e.g., measuring pairwise model agreement, analyzing information flow statistics, and testing multiple voting rules), we can compute \(R\) as a proxy for overall answer reliability in scenarios lacking ground truth. 

\paragraph{Methodological Overview}
To operationalize this framework, we conduct a set of experiments in which four leading LLMs each generate a suite of 100 PhD-level probabilistic questions. The remaining three models independently attempt to answer and justify each question. Roles rotate so that each model serves once as the “question generator” and thrice as an “answerer,” thus minimizing model-specific bias. Model outputs are then aggregated using both simple majority voting and more complex weighted consensus algorithms\footnote{Weighted consensus may incorporate confidence scores or token probabilities returned by the LLM API.}, and we evaluate convergence using metrics inspired by consensus formation theory \cite{mennis2006wisdom}.

\paragraph{Research Goals and Contributions}
Our investigation seeks to (1) identify how often and under what conditions multiple LLMs agree on probabilistic reasoning tasks; (2) determine which combinations of models yield the most reliable joint output; and (3) establish baseline benchmarks for collaborative LLM performance in the absence of ground truth. We anticipate that this work will make three primary contributions:
\begin{enumerate}
  \item A new multi-LLM validation framework that leverages collective intelligence, distributed cognition, and consensus principles to assess answer quality without relying on predefined correct responses.
  \item Empirical evidence demonstrating that collaborative validation can approach—or even surpass—individual-model accuracy in complex reasoning tasks.
  \item A set of benchmark agreement and reliability metrics to guide future research on large-scale LLM ensembles.
\end{enumerate}

\paragraph{Implications} 
The broader impact of this work spans several domains. In educational technology, our findings can inform automated assessment platforms that need to grade open-ended or highly technical questions. In research validation, collaborative LLM ensembles could accelerate literature reviews by cross-checking interpretations of novel statistical results. More generally, by demonstrating how multiple LLMs can validate each other’s outputs, we pave the way for AI-driven systems that maintain high reliability even as specialized knowledge evolves rapidly.

\section{Literature Review}
In recent years, LLMs and generative AI have become the primary focus, fundamentally altering our methods for tackling complex language-based tasks.


Modern language models excel at both comprehending and producing human language \cite{radford2019language}. The GPT series, for instance, delivers impressive few-shot and zero-shot performance \cite{achiam2023gpt}. Building upon this foundation, newer systems like Claude-3 \cite{anthropic2024claude3} and Gemini \cite{team2023gemini} extend these capabilities even further, demonstrating sophisticated reasoning in specialized areas.


A growing body of work has explored the advantages of collaborative intelligence, where multiple AI agents cooperate to solve problems. Studies show that combining outputs from different models enhances reasoning accuracy through cross-verification of responses \cite{yin2023exchange}. This multi-model interaction not only improves robustness when addressing complex tasks \cite{davoodi2024llms} but also produces more dependable results by leveraging consensus mechanisms.


Empirical evidence across diverse domains emphasize the benefits of ensemble and cooperative approaches. Lu \emph{et al.} \cite{lu2024merge} provide a comprehensive survey demonstrating that merging and ensembling techniques among LLMs yield better outcomes in various NLP tasks. In healthcare, Rezk \emph{et al.} \cite{rezk2024metaheuristic} review metaheuristic-driven ensemble learning methods and report improved diagnostic precision. An MIT-led initiative developed a multi-AI collaboration framework that enhanced accuracy and reasoning in complex mathematical problem-solving \cite{mit2023multi}. Similarly, ensemble strategies in software engineering have been shown to boost code defect detection and overall quality assurance processes \cite{gupta2022review}. These examples collectively highlight that collaborative AI approaches, as LLMs continue to evolve, can offer substantial performance gains.


However, coupling multiple models introduces important ethical considerations, particularly around transparency, accountability, and bias propagation \cite{bender2021dangers}. When ground truth is unavailable, guaranteeing that collaborative AI operates ethically becomes even more critical \cite{mittelstadt2019principles}. One central concern is the amplification of biases: Raghavan \emph{et al.} \cite{raghavan2020mitigating} demonstrate how ensemble methods can inadvertently magnify existing biases, resulting in disproportionate impacts on certain demographic groups. To address this, Hashimoto \emph{et al.} \cite{hashimoto2018fairness} propose frameworks designed to quantify and mitigate bias accumulation.


Ensuring transparency poses another challenge. Doshi-Velez and Kim \cite{doshi2017accountability} discuss the difficulty of tracing decisions and interpreting how consensus is achieved in multi-model systems, calling for stronger accountability structures. These issues become especially pressing in high-stakes domains where understanding the underlying reasoning is essential \cite{rudin2019stop}.


Several case studies further emphasize these ethical dimensions. In medical imaging, Larrazabal \emph{et al.} \cite{larrazabal2020gender} examine how collaborative systems raise privacy concerns and the need for transparent decision-making to protect patient interests. In education, Holstein \emph{et al.} \cite{holstein2019improving} explores issues of fairness in automated classification systems, illustrating how collaborative models must be carefully calibrated to ensure equitable evaluations. Gebru \emph{et al.} \cite{gebru2021datasheets} emphasized the importance of detailed documentation and transparency whenever multiple models are deployed together.


Overall, the literature indicates that while collaborative strategies among LLMs can significantly improve performance, they also necessitate careful consideration of ethical risks such as bias amplification and opacity. Addressing these challenges is crucial for developing trustworthy, scalable AI systems in specialized domains.

\section{Methodology}

This research adopts a mixed-methods strategy to explore collaborative interactions among LLMs in statistical reasoning tasks. The experimental design integrates quantitative assessments of consensus patterns across models. Formally, the study is defined as set of  M, Q, A, V where \(M\) denotes the set of models, \(Q\) represents the question generation process, \(A\) refers to the answering process, and \(V\) signifies the validation mechanisms.

The investigation centers on how multiple LLMs collaborate in generating and validating advanced probabilistic multiple-choice questions (MCQs) at the PhD level, absent ground-truth answers. In total, around \(N = 100\) MCQs were created and answered: one LLM generated each question while the remaining three independently provided answers and justifications. To reduce model-specific biases, we rotated roles among the models.

We utilized four leading LLMs—GPT-4, Meta-LLaMA-3-70B, Claude-3-Opus, and Gemini-1.5-Flash—each chosen for its distinct strengths: GPT-4 for advanced reasoning and extensive language comprehension; Meta-LLaMA-3-70B for its 70 billion-parameter scale and superior few-shot instruction-following; Claude-3-Opus for producing reliable, safety-focused outputs on nuanced queries; and Gemini-1.5-Flash for its multimodal integration of text and image understanding. All models were accessed via their official APIs with default hyperparameters and evaluated under the same conditions to ensure a fair comparison.

To generate challenging, diverse MCQs suitable for PhD-level probability, we developed a framework that combined question generation with independent answering. This framework also incorporates bias mitigation mechanisms.

The question-generation process was guided by a comprehensive concept map of advanced probabilistic topics. This map includes topics \(\mathcal{T}\) and subtopics \(\mathcal{S}\). For each MCQ, a topic \(t_i \in \mathcal{T}\) and subtopic \(s_i \in \mathcal{S}\) were randomly chosen. The assigned LLM received a structured prompt \(P_q\):
\begin{quote}
\textit{“Generate a challenging PhD-level multiple-choice question in the field of [\textbf{Topic}], focusing on [\textbf{Specific Concept}]. The question should have four answer options labeled A, B, C, and D, with only one correct answer. Ensure the question assesses deep understanding and critical thinking.”}
\end{quote}
Here, [\textbf{Topic}] and [\textbf{Specific Concept}] were filled according to \(t_i\) and \(s_i\). The output included the question \(Q_i\), answer set \(\{A_i, B_i, C_i, D_i\}\), the correct answer \(A_i^c\), and explanation \(E_i\). We stored \(A_i^c\) and \(E_i\) for later analysis but withheld them from answering models, simulating a real-world scenario without ground truth.

To mitigate bias, we implemented the following strategies:
\begin{enumerate}
  \item \textbf{Topic Diversity}: Ensured by including various topics and subtopics.
  \item \textbf{Neutral Prompts}: Crafted to avoid leading language or bias.
  \item \textbf{Manual Review}: All generated content was reviewed to exclude inappropriate or biased material.
\end{enumerate}

The study covered a broad spectrum of probability topics, including probability distributions (discrete distributions such as Binomial and Poisson, continuous distributions like Normal and Exponential); Bayesian probability (Bayes’ theorem, prior vs.\ posterior distributions, and likelihood functions); stochastic processes (Markov chains, Poisson processes, and random walks); Monte Carlo methods (sampling techniques, variance reduction strategies, and Monte Carlo integration); conditional probability (joint and marginal probabilities, independence and conditional independence, and the chain rule); information theory and entropy (Shannon entropy, mutual information, and conditional entropy); probability inequalities (Markov’s inequality, Chebyshev’s inequality, and the union bound); random variables (probability mass functions, probability density functions,  and transformations of random variables); limit theorems (the law of large numbers, and the central limit theorem ); and probabilistic graphical models (Bayesian networks, conditional independencies, inference algorithms in graphical models, and Markov random fields).

During the answering phase, three distinct LLMs independently tackled each generated question. Each received \(Q_i\) and \(\{A_i, B_i, C_i, D_i\}\) along with prompt \(P_a\):
\begin{quote}
\textit{“Please read the following PhD-level probabilistic question and select the most appropriate answer (A, B, C, or D). Provide a detailed justification for your selection, explaining your reasoning and any relevant statistical principles.”}
\end{quote}
Each model returned an answer \(a_{ij}\) (where \(j \in \{1,2,3\}\)) and a justification \(J_{ij}\). Models were isolated to prevent information exchange, and neutral prompts plus diverse topics further minimized bias.

We performed inter-model consistency analysis to assess agreement and reliability. For each question \(Q_i\), we constructed a dataset:
\begin{equation}
D_i = \{Q_i,\ \{A_i,B_i,C_i,D_i\},\ \{a_{i1}, a_{i2}, a_{i3}\},\ \{J_{i1},J_{i2},J_{i3}\}\}.
\end{equation}
This dataset included \(Q_i\), the multiple-choice options, each model’s selected answer \(a_{ij}\), justifications \(J_{ij}\), and metadata. This structure supported quantitative and qualitative analyses of reasoning and agreement patterns \cite{bommasani2021opportunities}.

We divided model responses into three categories: complete agreement, when all three models chose the same answer; partial agreement, when two models agreed while the third differed; and no agreement, when each model selected a different option. This scheme allowed us to quantify overall consensus and examine how responses varied across models.


\subsection{Consensus and Reliability}
Without ground truth, we validated responses based on inter-model consensus.

\subsubsection{Majority Vote}
To determine the consensus answer \(A_i^{\text{cons}}\) for each \(Q_i\), we used a majority voting system:
\begin{equation}
A_i^{\text{cons}} = \arg\max_{k \in \{A,B,C,D\}} \sum_{j=1}^{3} \delta(a_{ij}, k),
\end{equation}
where \(\delta(a_{ij}, k)\) is 1 if \(a_{ij} = k\), and 0 otherwise. The answer with highest frequency was taken as consensus.

\subsubsection{Reliability Metric}
We introduced a reliability score \(R_i\) to assess trustworthiness:
\begin{equation}
R_i =
\begin{cases}
1, & \text{if }A_i^{\text{cons}} = A_i^{\text{LLM-q}},\\
0, & \text{otherwise,}
\end{cases}
\end{equation}
where \(A_i^{\text{LLM-q}}\) is the querying LLM’s own answer. A higher \(R_i\) indicates stronger alignment between consensus and the question-generator’s answer.

\subsubsection{Confidence Intervals via Bootstrap}
To evaluate robustness of consensus rates, we computed 95\% confidence intervals (CIs) using bootstrap resampling. We generated \(B\) bootstrap samples by sampling with replacement from the original dataset. For each sample, we calculated the mean agreement rate, forming a bootstrap distribution. The 2.5th and 97.5th percentiles provided the lower and upper bounds of the CI. Narrow CIs imply high reliability; non-overlapping CIs suggest significant differences among models.

\subsection{Statistical Tests}
\subsubsection{Chi-Square Test of Independence}
We applied a chi-square test to determine whether the distribution of selected answers deviated from uniform random selection. The test statistic is:
\begin{equation}
\chi^2 = \sum_{k=1}^{K} \frac{(O_k - E_k)^2}{E_k},
\label{chi_square_test}
\end{equation}
where:
\begin{itemize}
  \item \(O_k\) is the observed frequency of answer choice \(k\).
  \item \(E_k\) is the expected frequency under uniform selection: \(E_k = \frac{N \times n_j}{K}\).
  \item \(K = 4\) (number of answer choices), \(N\) is total questions, and \(n_j = 3\) is number of answering models.
\end{itemize}
The computed \(\chi^2\) is compared to the chi-square distribution with \(K - 1\) degrees of freedom.

\subsubsection{Fleiss’ Kappa}
Fleiss’ kappa \(\kappa\) measures agreement among models beyond chance:
\begin{equation}
\kappa = \frac{\overline{P} - \overline{P_e}}{1 - \overline{P_e}},
\end{equation}
where \(\overline{P}\) is the average observed agreement and \(\overline{P_e}\) is the average chance agreement. \(\kappa\) ranges from \(-1\) (complete disagreement) to \(1\) (perfect agreement), with \(0\) indicating no agreement beyond chance.

Combined, these methods offer a rigorous framework to evaluate model consensus and reliability in the absence of ground truth.

\section{Results}

We evaluated how consistently the four LLMs (GPT-4, LLaMA, Gemini, and Claude) agreed when answering 100 questions generated by each model in turn. Table~\ref{tab:consensus_rates_small} shows the proportions of full, partial, and no agreement across all experiments. When Gemini or Claude created the questions, the other models fully agreed on the same answer over 73\% of the time—74.0\% for Gemini-generated items and 73.1\% for Claude-generated items. GPT-4’s question set yielded full agreement 64.7\% of the time, while LLaMA’s questions saw only 49.5\% full agreement. LLaMA also produced the highest rate of partial agreement (40.9\%) and the largest “no agreement” rate (9.7\%), indicating its prompts introduced more uncertainty. In contrast, Claude had the fewest complete disagreements (1.1\%) and the lowest partial agreement (25.8\%).

\begin{table}[h]
  \centering
  \scriptsize
  \caption{Consensus rates by question-generating model}
  \label{tab:consensus_rates_small}
  \begin{tabular}{lccc}
    \toprule
    \textbf{Model} & \textbf{Full (\%)} & \textbf{Partial (\%)} & \textbf{None (\%)} \\
    \midrule
    Gemini & 74.0 & 22.0 & 4.0 \\
    Claude & 73.1 & 25.8 & 1.1 \\
    GPT-4  & 64.7 & 26.5 & 8.8 \\
    LLaMA  & 49.5 & 40.9 & 9.7 \\
    \bottomrule
  \end{tabular}
\end{table}

To measure reliability without ground truth, we applied a majority‐vote rule and measured how often the consensus matched the question‐generator’s original answer. Table~\ref{tab:majority_reliability} reports both the percentage of cases where at least two models agreed (“Majority Vote”) and the percentage where that majority also agreed with the generating model’s answer (“Reliability”). Claude questions achieved a unanimous majority in all trials (100\%), with 88.2\% of these consensus answers matching the original. GPT-4 and Gemini similarly showed high reliability (87.3\% and 79.0\%, respectively) and majority‐vote rates above 90\%. LLaMA lagged behind, with 90.3\% majority‐vote consistency but only 66.7\% reliability, reflecting that its question set often steered other models toward different answers than its own.

\begin{table}[h]
  \centering
  \caption{Majority-vote consistency and alignment with the generating model.}
  \label{tab:majority_reliability}
  \begin{tabular}{lcc}
    \toprule
    \textbf{Model} & \textbf{Majority Vote (\%)} & \textbf{Reliability (\%)} \\
    \midrule
    Gemini & 96.0 & 79.0 \\
    Claude & 100.0 & 88.2 \\
    GPT-4 & 91.2 & 87.3 \\
    LLaMA & 90.3 & 66.7 \\
    \bottomrule
  \end{tabular}
\end{table}

Next, we calculated 95\% bootstrap confidence intervals (CIs) for each model’s full‐agreement rate (Table~\ref{tab:ci}). Claude displayed the narrowest CI (0.70–0.86), implying consistent high agreement, followed closely by GPT-4 (0.63–0.80). Gemini’s interval was slightly wider (0.59–0.78), whereas LLaMA showed both the lowest and most variable rates (0.29–0.49), underscoring its lower reliability and greater fluctuation in consensus.

\begin{table}[h]
  \centering
  \caption{Bootstrap 95\% confidence intervals for full‐agreement rates.}
  \label{tab:ci}
  \begin{tabular}{lccc}
    \toprule
    \textbf{Model} & \textbf{Lower} & \textbf{Upper} & \textbf{Width} \\
    \midrule
    Gemini & 0.59 & 0.78 & 0.19 \\
    Claude & 0.70 & 0.86 & 0.16 \\
    GPT-4  & 0.63 & 0.80 & 0.17 \\
    LLaMA  & 0.29 & 0.49 & 0.20 \\
    \bottomrule
  \end{tabular}
\end{table}

To verify that these agreement levels were not due to chance, we performed chi‐square tests (Table~\ref{tab:chi_square}). For Gemini, Claude, and GPT-4, the very small $p$‐values (\(<10^{-5}\)) reject the null hypothesis of random agreement, confirming statistically significant consensus. LLaMA’s $p$‐value (5.66 × 10\(^{-2}\)) did not pass the 0.01 threshold, indicating its observed agreement could plausibly arise by chance more often.

\begin{table}[h]
  \centering
  \caption{Chi‐square test $p$‐values (significance threshold: 0.01).}
  \label{tab:chi_square}
  \begin{tabular}{lcccc}
    \toprule
    \textbf{Model} & \textbf{Gemini} & \textbf{Claude} & \textbf{GPT-4} & \textbf{LLaMA} \\
    \midrule
    \textbf{$p$‐value} & \(1.95\times10^{-16}\) & \(4.34\times10^{-32}\) & \(2.65\times10^{-5}\) & \(5.66\times10^{-2}\) \\
    \bottomrule
  \end{tabular}
\end{table}

Finally, we computed Fleiss’ kappa to quantify inter‐model agreement beyond chance (Table~\ref{tab:kappa}). Gemini achieved the highest kappa (0.622, “substantial” agreement), followed by Claude at 0.520 (“moderate”). GPT-4 and LLaMA both fell into the “fair” category (0.387 and 0.279, respectively), highlighting greater variability in their answers. These measures confirm that, while Gemini and Claude produce highly consistent and reliable question sets, GPT-4 performs moderately well, and LLaMA’s question clarity and resulting consensus require further improvement.

\begin{table}[h]
  \centering
  \caption{Fleiss’ kappa values and agreement interpretations.}
  \label{tab:kappa}
  \begin{tabular}{lcc}
    \toprule
    \textbf{Model} & \textbf{Kappa} & \textbf{Interpretation} \\
    \midrule
    Gemini & 0.622 & Substantial agreement \\
    Claude & 0.520 & Moderate agreement \\
    GPT-4  & 0.387 & Fair agreement \\
    LLaMA  & 0.279 & Fair agreement \\
    \bottomrule
  \end{tabular}
\end{table}





\section{Conclusion}

We examined how multiple LLMs collaborate to create and validate advanced probabilistic questions without relying on ground truth. By measuring inter-model agreement, reliability, and consensus, we uncovered notable differences in question clarity and interpretability across models.

Claude and GPT-4 consistently produced well-structured prompts, yielding high agreement rates and tight 95\% confidence intervals (Claude: 0.70–0.86; GPT-4: 0.63–0.80). In contrast, Gemini (0.59–0.78) and LLaMA (0.29–0.49) showed wider intervals, reflecting greater variability in how other models interpreted their questions. Chi-square tests and Fleiss’ kappa confirmed that most agreements were statistically significant rather than random.

These results underscore the importance of question quality: models that generate clear, unambiguous prompts foster stronger consensus in multi-LLM settings. They also highlight LLaMA’s need for further refinement, given its lower reliability (66.7\%) and higher disagreement rates. Overall, majority voting and confidence-interval analysis emerged as practical methods for approximating answer validity when no ground truth exists. Looking ahead, incorporating expert evaluations could benchmark these consensus measures against human-verified answers. Additionally, studying how biases propagate through inter-model agreements will help prevent systematic errors. Addressing these aspects  enhances robustness and trustworthiness of collaborative LLM-based validation systems.

\bibliographystyle{IEEEtran}
\bibliography{ref}

\begin{thebibliography}{10}
\providecommand{\url}[1]{#1}
\csname url@samestyle\endcsname
\providecommand{\newblock}{\relax}
\providecommand{\bibinfo}[2]{#2}
\providecommand{\BIBentrySTDinterwordspacing}{\spaceskip=0pt\relax}
\providecommand{\BIBentryALTinterwordstretchfactor}{4}
\providecommand{\BIBentryALTinterwordspacing}{\spaceskip=\fontdimen2\font plus
\BIBentryALTinterwordstretchfactor\fontdimen3\font minus \fontdimen4\font\relax}
\providecommand{\BIBforeignlanguage}[2]{{%
\expandafter\ifx\csname l@#1\endcsname\relax
\typeout{** WARNING: IEEEtran.bst: No hyphenation pattern has been}%
\typeout{** loaded for the language `#1'. Using the pattern for}%
\typeout{** the default language instead.}%
\else
\language=\csname l@#1\endcsname
\fi
#2}}
\providecommand{\BIBdecl}{\relax}
\BIBdecl

\bibitem{guo2025deepseek}
D.~Guo, D.~Yang, H.~Zhang, J.~Song, R.~Zhang, R.~Xu, Q.~Zhu, S.~Ma, P.~Wang, X.~Bi \emph{et~al.}, ``Deepseek-r1: Incentivizing reasoning capability in llms via reinforcement learning,'' \emph{arXiv preprint arXiv:2501.12948}, 2025.

\bibitem{touvron2023LLaMA}
H.~Touvron, T.~Lavril, G.~Izacard, X.~Martinet, M.-A. Lachaux, T.~Lacroix, B.~Rozi{\`e}re, N.~Goyal, E.~Hambro, F.~Azhar \emph{et~al.}, ``Llama: Open and efficient foundation language models,'' \emph{arXiv preprint arXiv:2302.13971}, 2023.

\bibitem{kojima2022large}
T.~Kojima, S.~S. Gu, M.~Reid, Y.~Matsuo, and Y.~Iwasawa, ``Large language models are zero-shot reasoners,'' \emph{Advances in neural information processing systems}, vol.~35, pp. 22\,199--22\,213, 2022.

\bibitem{amiri2024enhancing}
A.~Amiri-Margavi, I.~Jebellat, E.~Jebellat, and S.~P.~M. Davoudi, ``Enhancing answer reliability through inter-model consensus of large language models,'' \emph{arXiv preprint arXiv:2411.16797}, 2024.

\bibitem{hendrycks2020measuring}
D.~Hendrycks, C.~Burns, S.~Basart, A.~Zou, M.~Mazeika, D.~Song, and J.~Steinhardt, ``Measuring massive multitask language understanding,'' \emph{arXiv preprint arXiv:2009.03300}, 2020.

\bibitem{he2022wisdom}
L.~He, P.~P. Analytis, and S.~Bhatia, ``The wisdom of model crowds,'' \emph{Management Science}, vol.~68, no.~5, pp. 3635--3659, 2022.

\bibitem{dietterich2000ensemble}
T.~G. Dietterich, ``Ensemble methods in machine learning,'' in \emph{International workshop on multiple classifier systems}.\hskip 1em plus 0.5em minus 0.4em\relax Springer, 2000, pp. 1--15.

\bibitem{mennis2006wisdom}
E.~A. Mennis, ``The wisdom of crowds: Why the many are smarter than the few and how collective wisdom shapes business, economies, societies, and nations,'' \emph{Business Economics}, vol.~41, no.~4, pp. 63--65, 2006.

\bibitem{taylor2022galactica}
R.~Taylor, M.~Kardas, G.~Cucurull, T.~Scialom, A.~Hartshorn, E.~Saravia, A.~Poulton, V.~Kerkez, and R.~Stojnic, ``Galactica: A large language model for science,'' \emph{arXiv preprint arXiv:2211.09085}, 2022.

\bibitem{ahn2024large}
J.~Ahn, R.~Verma, R.~Lou, D.~Liu, R.~Zhang, and W.~Yin, ``Large language models for mathematical reasoning: Progresses and challenges,'' \emph{arXiv preprint arXiv:2402.00157}, 2024.

\bibitem{huang2024enabling}
Y.~Huang, X.~Feng, B.~Li, Y.~Xiang, H.~Wang, B.~Qin, and T.~Liu, ``Enabling ensemble learning for heterogeneous large language models with deep parallel collaboration,'' \emph{arXiv preprint arXiv:2404.12715}, 2024.

\bibitem{patrikalakis1999distributed}
N.~M. Patrikalakis, P.~J. Fortier, Y.~Ioannidis, C.~N. Nikolaou, A.~R. Robinson, J.~R. Rossignac, A.~Vinacua, and S.~L. Abrams, ``Distributed information and computation in scientific and engineering environments,'' \emph{D-lib Magazine}, vol.~5, no.~4, pp. 1082--9873, 1999.

\bibitem{levy1997collective}
P.~L{\'e}vy, \emph{Collective intelligence: Mankind's emerging world in cyberspace}.\hskip 1em plus 0.5em minus 0.4em\relax Perseus books, 1997.

\bibitem{woolley2010evidence}
A.~W. Woolley, C.~F. Chabris, A.~Pentland, N.~Hashmi, and T.~W. Malone, ``Evidence for a collective intelligence factor in the performance of human groups,'' \emph{science}, vol. 330, no. 6004, pp. 686--688, 2010.

\bibitem{hutchins1995cognition}
E.~Hutchins, \emph{Cognition in the wild}.\hskip 1em plus 0.5em minus 0.4em\relax MIT Press, 1995.

\bibitem{du2024large}
Y.~Du, P.~Rajivan, and C.~Gonzalez, ``Large language models for collective problem-solving: Insights into group consensus,'' in \emph{Proceedings of the Annual Meeting of the Cognitive Science Society, 46 (0)}, 2024.

\bibitem{naik2024probabilistic}
N.~Naik, ``Probabilistic consensus through ensemble validation: A framework for llm reliability,'' \emph{arXiv preprint arXiv:2411.06535}, 2024.

\bibitem{friedkin1990social}
N.~E. Friedkin, ``Social networks in structural equation models,'' \emph{Social Psychology Quarterly}, pp. 316--328, 1990.

\bibitem{olfati2007consensus}
R.~Olfati-Saber, J.~A. Fax, and R.~M. Murray, ``Consensus and cooperation in networked multi-agent systems,'' \emph{Proceedings of the IEEE}, vol.~95, no.~1, pp. 215--233, 2007.

\bibitem{baronchelli2018emergence}
A.~Baronchelli, ``The emergence of consensus: a primer,'' \emph{Royal Society open science}, vol.~5, no.~2, p. 172189, 2018.

\bibitem{bahrami2010optimally}
B.~Bahrami, K.~Olsen, P.~E. Latham, A.~Roepstorff, G.~Rees, and C.~D. Frith, ``Optimally interacting minds,'' \emph{Science}, vol. 329, no. 5995, pp. 1081--1085, 2010.

\bibitem{radford2019language}
A.~Radford, J.~Wu, R.~Child, D.~Luan, D.~Amodei, I.~Sutskever \emph{et~al.}, ``Language models are unsupervised multitask learners,'' \emph{OpenAI blog}, vol.~1, no.~8, p.~9, 2019.

\bibitem{achiam2023gpt}
J.~Achiam, S.~Adler, S.~Agarwal, L.~Ahmad, I.~Akkaya, F.~L. Aleman, D.~Almeida, J.~Altenschmidt, S.~Altman, S.~Anadkat \emph{et~al.}, ``Gpt-4 technical report,'' \emph{arXiv preprint arXiv:2303.08774}, 2023.

\bibitem{anthropic2024claude3}
Anthropic, ``The claude 3 model family: Opus, sonnet, haiku,'' \emph{Anthropic Technical Report}, 2024.

\bibitem{team2023gemini}
G.~Team, R.~Anil, S.~Borgeaud, J.-B. Alayrac, J.~Yu, R.~Soricut, J.~Schalkwyk, A.~M. Dai, A.~Hauth, K.~Millican \emph{et~al.}, ``Gemini: a family of highly capable multimodal models,'' \emph{arXiv preprint arXiv:2312.11805}, 2023.

\bibitem{yin2023exchange}
Z.~Yin, Q.~Sun, C.~Chang, Q.~Guo, J.~Dai, X.~Huang, and X.~Qiu, ``Exchange-of-thought: Enhancing large language model capabilities through cross-model communication,'' \emph{arXiv preprint arXiv:2312.01823}, 2023.

\bibitem{davoodi2024llms}
A.~G. Davoodi, S.~P.~M. Davoudi, and P.~Pezeshkpour, ``Llms are not intelligent thinkers: Introducing mathematical topic tree benchmark for comprehensive evaluation of llms,'' \emph{arXiv preprint arXiv:2406.05194}, 2024.

\bibitem{lu2024merge}
J.~Lu, Z.~Pang, M.~Xiao, Y.~Zhu, R.~Xia, and J.~Zhang, ``Merge, ensemble, and cooperate! a survey on collaborative strategies in the era of large language models,'' \emph{arXiv preprint arXiv:2407.06089}, 2024.

\bibitem{rezk2024metaheuristic}
S.~S. Rezk and K.~S. Selim, ``Metaheuristic-based ensemble learning: an extensive review of methods and applications,'' \emph{Neural Computing and Applications}, vol.~36, no.~29, pp. 17\,931--17\,959, 2024.

\bibitem{mit2023multi}
\BIBentryALTinterwordspacing
{MIT News}, ``Multi-ai collaboration helps reasoning and factual accuracy in large language models,'' 2023. [Online]. Available: \url{https://news.mit.edu/2023/multi-ai-collaboration-helps-reasoning-factual-accuracy-language-models-0918}
\BIBentrySTDinterwordspacing

\bibitem{gupta2022review}
P.~Gupta, A.~Pratap~Singh, and V.~Kumar, ``A review of ensemble methods used in ai applications,'' in \emph{International Conference on Cybersecurity in Emerging Digital Era}.\hskip 1em plus 0.5em minus 0.4em\relax Springer, 2022, pp. 145--157.

\bibitem{bender2021dangers}
E.~M. Bender, T.~Gebru, A.~McMillan-Major, and S.~Shmitchell, ``On the dangers of stochastic parrots: Can language models be too big?'' in \emph{Proceedings of the 2021 ACM conference on fairness, accountability, and transparency}, 2021, pp. 610--623.

\bibitem{mittelstadt2019principles}
B.~Mittelstadt, ``Principles alone cannot guarantee ethical ai,'' \emph{Nature machine intelligence}, vol.~1, no.~11, pp. 501--507, 2019.

\bibitem{raghavan2020mitigating}
M.~Raghavan, S.~Barocas, J.~Kleinberg, and K.~Levy, ``Mitigating bias in algorithmic hiring: Evaluating claims and practices,'' in \emph{Proceedings of the 2020 conference on fairness, accountability, and transparency}, 2020, pp. 469--481.

\bibitem{hashimoto2018fairness}
T.~Hashimoto, M.~Srivastava, H.~Namkoong, and P.~Liang, ``Fairness without demographics in repeated loss minimization,'' in \emph{International Conference on Machine Learning}.\hskip 1em plus 0.5em minus 0.4em\relax PMLR, 2018, pp. 1929--1938.

\bibitem{doshi2017accountability}
F.~Doshi-Velez, M.~Kortz, R.~Budish, C.~Bavitz, S.~Gershman, D.~O'Brien, K.~Scott, S.~Schieber, J.~Waldo, D.~Weinberger \emph{et~al.}, ``Accountability of ai under the law: The role of explanation,'' \emph{arXiv preprint arXiv:1711.01134}, 2017.

\bibitem{rudin2019stop}
C.~Rudin, ``Stop explaining black box machine learning models for high stakes decisions and use interpretable models instead,'' \emph{Nature machine intelligence}, vol.~1, no.~5, pp. 206--215, 2019.

\bibitem{larrazabal2020gender}
A.~J. Larrazabal, N.~Nieto, V.~Peterson, D.~H. Milone, and E.~Ferrante, ``Gender imbalance in medical imaging datasets produces biased classifiers for computer-aided diagnosis,'' \emph{Proceedings of the National Academy of Sciences}, vol. 117, no.~23, pp. 12\,592--12\,594, 2020.

\bibitem{holstein2019improving}
K.~Holstein, J.~Wortman~Vaughan, H.~Daum{\'e}~III, M.~Dudik, and H.~Wallach, ``Improving fairness in machine learning systems: What do industry practitioners need?'' in \emph{Proceedings of the 2019 CHI conference on human factors in computing systems}, 2019, pp. 1--16.

\bibitem{gebru2021datasheets}
T.~Gebru, J.~Morgenstern, B.~Vecchione, J.~W. Vaughan, H.~Wallach, H.~D. Iii, and K.~Crawford, ``Datasheets for datasets,'' \emph{Communications of the ACM}, vol.~64, no.~12, pp. 86--92, 2021.

\bibitem{bommasani2021opportunities}
Bommasani, ``On the opportunities and risks of foundation models,'' \emph{arXiv preprint arXiv:2108.07258}, 2021.

\end{thebibliography}


\end{document}